\documentstyle[prd,aps,floats]{revtex}
\newcommand{\be}{\begin{equation}}
\newcommand{\ee}{\end{equation}}
\newcommand{\bea}{\begin{eqnarray}}
\newcommand{\beas}{\begin{eqnarray*}}
\newcommand{\eea}{\end{eqnarray}}
\newcommand{\eeas}{\end{eqnarray*}}
\newcommand{\ba}{\begin{array}}
\newcommand{\ea}{\end{array}}
\newcommand{\bi}{\begin{itemize}}
\newcommand{\ei}{\end{itemize}}
\newcommand{\ben}{\begin{enumerate}}
\newcommand{\een}{\end{enumerate}}
\begin{document}
\draft

\input epsf \renewcommand{\topfraction}{0.8} 
\twocolumn[\hsize\textwidth\columnwidth\hsize\csname 
@twocolumnfalse\endcsname

%preprint No.DFAQ-2001/04-TH 

\title{Affleck-Dine leptogenesis via right-handed sneutrino fields\\
 in a supersymmetric  hybrid inflation model} 
\author{Zurab Berezhiani$^{a,b}$, Anupam  Mazumdar$^c$ and 
Abdel P\'erez-Lorenzana$^{c,d}$}
\address{$^a$Dipartamento di Fisica, Universit\'a di L'Aquila, I-67010, 
Coppito, AQ, and, \\
INFN, Laboratori Nazionali del Gran Sasso, I-67010, Assergi, AQ, Italy \\
$^b$ The Andronikashvili Institute of Physics, GE-380077 Tbilisi, Georgia
\\
$^c$ The Abdus Salam International Centre for Theoretical Physics, I-34100,
Trieste, Italy \\
$^d$ Departamento de F\'{\i}sica,
Centro de Investigaci\'on y de Estudios Avanzados del I.P.N.\\
Apdo. Post. 14-740, 07000, M\'exico, D.F., M\'exico}
\date{\today} 
\maketitle

\begin{abstract}
The onset of inflation in hybrid models require fine tuning in 
the initial conditions. The inflaton field should have an initial
value close to the Planck scale $M_{\rm P}$, whereas the auxiliary "orthogonal"
field must be close to zero with an extreme accuracy. 
This problem can be alleviated if the orthogonal field fastly 
decays into some states not coupled to the inflaton. 
Natural candidates for such states can be the right-handed neutrinos. 
We show that a non-trivial evolution of the classic sneutrino 
fields after inflation offers 
an interesting mechanism for generating a correct amount 
of lepton asymmetry, which being reprocessed by sphalerons 
can explain the observed  baryon asymmetry of the Universe. 
Our scenario implies interesting bounds for the neutrino 
masses in the context of seesaw mechanism.
\end{abstract}

\vskip2pc]

%\vspace*{-1.0truecm}

%%%%%%%%%%%%%%%%%%%%%%%%%%%%%%%%%%%%%%%%%%%%%%%%%%%%%%%%%%%%%%%%%%%%%%%%
\section{introduction} 

The hybrid inflation model \cite{linde0} provides an attractive 
possibility for solving a range of cosmological problems \cite{liddle}.  
It has an unique dynamical feature due to interplay between two 
scalar fields, inflaton $\sigma$ and an auxiliary field,  
so called orthogonal scalar $\phi$ which is actually responsible 
for providing the potential energy throughout the inflationary phase.
Quite remarkably, a proper form of the inflaton potential 
can be naturally realized in the context of supersymmetric 
theories \cite{susy1,susy2}. 
The simplest supersymmetric model for hybrid inflation  
is based on the following superpotential 
including the inflaton superfield $S$ and
the auxiliary one $\Phi$ \cite{susy1}:  
\begin{equation} \label{super}
W=-\Lambda^2 S +\lambda S \Phi^2 \,,
\end{equation}
where $\Lambda$ and $\lambda$ are the model parameters. 
The latter are constrained by COBE normalization for the density 
perturbations as 
$\Lambda\approx 6.5 \cdot 10^{16}\epsilon^{1/4}$ GeV \cite{liddle},  
or equivalently  
\begin{equation} \label{eta}
\Lambda \approx 1.3 \cdot  10^{15} |\eta|\lambda^{-1/2}~ {\rm GeV}\, ,
\end{equation}
where $\epsilon,\eta \ll 1$ are the slow roll parameters.  

The scalar components of these superfields 
have a potential 
\begin{equation}
\label{pot0}
V(\sigma,\phi)
=\Lambda^4-\lambda\Lambda^2\phi^2+\frac{\lambda^2}{4}\phi^4
+\lambda^2\sigma^2\phi^2 \,. 
\end{equation}
which has a global (supersymmetric) minima at 
$\sigma=0$ and $\phi^2=2\Lambda^2/\lambda$. 
However, for $\phi=0$ and 
$\sigma > \sigma_{c}=\Lambda/\sqrt{\lambda}$,  
the potential is flat along the $\sigma$ axis  
with non-zero vacuum energy $V(\sigma,0) =\Lambda^4$.    
This flat direction is lifted by radiative corrections resulting 
from the supersymmetry breaking and perhaps also by some  
supergravity corrections \cite{susy1}   
that give an appropriate slope to the inflaton 
potential on which $\sigma$ can slowly roll down and 
produce inflation. During inflation the orthogonal field 
is held in a false vacuum $\phi=0$, while the 
inflaton must evolve from an initial value 
close to the Planck scale $M_{\rm P} = 1.2 \times 10^{19}$ GeV.  
The inflationary phase ends when the inflaton field approaches 
the critical value $\sigma_c$, after which both fields 
$\sigma$ and $\phi$ begin to oscillate near their global minima 
and reheat the universe.

However, as any inflationary paradigm, hybrid inflation
models also suffer from the initial condition problems. 
A generic problem concerns the difficulty to inflate 
an arbitrary space-time patch.   
In order to trigger inflation, 
the inflaton field should be extremely homogeneous
in an initial patch of the Universe at scales larger 
than the Hubble radius of this epoch \cite{piran}.
This becomes a challenging task, since we know from the 
observations that the present inhomogeneity 
is primordial in nature and necessarily ties its origin 
with inflation. This obstacle can be  evaded for a chaotic 
initial condition provided the fields take values close to 
the Planck  scale.

In the case of hybrid inflation this  generic problem is 
aggravated by the fact that strong fine tuning is required 
also for the initial values for the fields \cite{tetradis,zurab}. 
This consists in the following. At the Planck epoch, one can expect 
that  all scalar fields, not only the inflaton,  take their initial values 
close to the $M_{\rm P}$. However, if the orthogonal field has a value 
$\phi\sim M_{\rm P}$, it does not provide the right condition for 
the onset of inflation. The reason is simply the following. 
While the field $\sigma\sim M_{\rm P}$ provides large 
effective mass term to $\phi$ through the last term 
in ~(\ref{pot0}), the value $\phi \sim M_{\rm P}$ in turn would induce 
large mass term to $\sigma$. In this case both the auxiliary field and 
inflaton would merely oscillate at their local minima without 
inflating the Universe. In order to trigger inflation, 
the initial field configuration should be settled down
along the $\sigma$-valley, with $\sigma$ having the value of 
order $M_P$ while $\phi$ must be close to zero with extreme 
accuracy, $\phi < 10^{-5} M_{\rm P}$ \cite{tetradis,zurab}. 
This is precisely the point which is not very natural; 
why the orthogonal field  should start so close to the false minimum 
of its potential? In other terms, in order to inflate the initial 
patch of space after the Planck epoch, it should not only be strongly 
homogeneous at distances much larger than the corresponding Hubble radius, 
but it also should have energy density $\sim \Lambda^4$, 
about 15 orders of magnitude smaller than $M_{\rm P}$.

Certainly, such stringent initial conditions can be 
accepted on purely anthropic grounds. However, it is always 
appealing to obtain a more natural solution of the problem. 
It has been shown recently in Ref.~\cite{zurab}, 
that the supersymmetric hybrid model can allow a
solution to this fine-tuning  problem quite elegantly. 
The solution is simple as it assumes that all the fields  take 
their initial value close to the Planck scale. 
However, the orthogonal field $\phi$ must decay into the 
quanta of some extra fields which do not interact with 
the inflaton, and thereby do not contribute to the curvature 
of the inflaton potential. This can be easily achieved by adding to 
the superpotential~(\ref{super}) a term $\kappa\Phi\Psi^2$, 
where $\Psi$ is the extra superfield.  
The  field $\phi$ oscillating near its false minimum, which one 
can consider as a matter dominated phase before
the onset of inflation, decays into $\Psi$ and settles down to zero
quickly enough such that the value of the inflaton remains 
$\sigma\sim M_{\rm P}$ which allows inflation to begin.

The graceful exit of inflation take place when the inflaton 
crosses the critical point on the valley which unfolds the 
strong positive curvature of the false vacuum of an auxiliary 
field to a smooth potential with a negative curvature which 
allows this field to roll down to its true minimum.
When this happens both the inflaton and the auxiliary field
begin oscillations around their respective global minima. 
One of the interesting aftermath of any inflationary dynamics 
is that the fields which take part in inflating the Universe, 
produce entropy and a thermal bath during their oscillations, 
the process known as the reheating of the Universe.

Shortly after, or possibly during reheating another crucial 
event, must occur: baryogenesis. As far as any preexisting 
baryon asymmetry has been exponentially diluted during 
inflation, one has to find a valid mechanism for 
generating the observed baryon asymmetry.

By now there are plenty of models for baryogenesis. 
One of the attractive possibilities would be to produce the 
baryon asymmetry just during the reheating, 
in an out of equilibrium decay of the inflaton particles 
themselves. 
Notice, that the departure from the thermal equilibrium 
is one of the three basic criterion for baryogenesis
other than $CP$ and $B$ (or rather $B-L$) violation \cite{sakh}. 
It is not easy, however, to naturally realize this 
possibility in the context of supersymmetric hybrid inflation.

Another interesting idea is related to leptogenesis 
\cite{buchmuller}, where first 
the non-zero lepton number $L$ is generated in some 
out-of-equilibrium processes, usually assumed as  
a right-handed neutrino \cite{fukugita},
and it is partially reprocessed into baryon number $B$ via 
$B+L$ violating sphaleron effects which however preserve
$B-L$ \cite{thooft}.

Yet another approach, known as Affleck-Dine mechanism \cite{affleck},
takes advantage of the flat directions  in supersymmetry 
which carry $B-L$ global charge. The non-zero baryon number 
density can be produced in decay of these modes at later times 
in the Universe evolution.

Baryogenesis mechanism which we  propose in this paper 
is a blend of these three ideas discussed above, and 
it can be naturally realized in the context of 
the supersymmetric hybrid inflation model \cite{zurab}.

Namely, a new step which we make here is to  
identify the extra superfield $\Psi$, needed for 
alleviating the problem of initial conditions,  
as a right-handed (RH) neutrino. 
First of all, this proposal can be motivated 
by the observation that the inflationary energy scale 
$\Lambda \sim 10^{15}$ GeV is also of interest as the RH neutrino 
mass scale in the context of the see saw mechanism \cite{seesaw}. 
Indeed, in the global minimum the field $\phi$ receives 
a vacuum expectation value (VEV) and thus  
the coupling $\Phi\Psi^2$ induces the mass 
$M_\Psi \sim \Lambda$.

Second, in this case is the scalar component 
of this superfield, the RH sneutrino field $\tilde\Phi$, 
carries the lepton number. 
The associated exact global symmetry, which is actually 
$U(1)_{B-L}$, is violated by the VEV of $\phi$. 
However, during the inflation the system is trapped in a false vacuum 
with $\phi=0$, where the $U(1)_{B-L}$ symmetry is restored 
and $\tilde\Psi$ behaves as massless field.  
Along with $\sigma$ and $\phi$,  also $\tilde\Psi$ must 
have initial value $\sim M_{\rm P}$ and before inflation 
it oscillates around zero. However, once the field $\phi$ 
decays, $\tilde\Psi$ become essentially massless mode. 
Therefore, during the de Sitter phase its evolution is slow 
and finally at the end of inflation it has  
non-zero value of the order of the Hubble parameter 
$H\sim \Lambda^2/M_{\rm P}$.  
In postinflationary epoch this  field becomes 
massive and starts to oscillate near the origin. 
As we show below, their evolution in this epoch 
epoch could generate, due to  dynamical breaking of 
the associated global $U(1)$ symmetry, an adequate amount 
of $B-L$ in the Universe, which can be converted into 
the baryon asymmetry via sphaleron transitions.

The paper is organized as follows. We begin with presenting 
our model. In next section we discuss the evolution patterns 
of the scalar fields before, during, and, after inflation 
and the reheating of the Universe. Then we calculate the 
lepton asymmetry generated by the RH sneutrino field and 
discuss see-saw  phenomenology for neutrino masses. 
Finally, we conclude with a brief discussion of our findings.

%%%%%%%%%%%%%%%%%%%%%%%%%%%%%%%%%%%%%%%%%%%%%%%%%%%%%%%%%%%%%%%%%%%%%%%%%%

\section{The model}

From the point of view of the particle physics, our model 
is nothing but the MSSM including the standard fermion superfields:  
leptons $l=(\nu,e)$, $e^c$ and quarks 
(which we do not write explicitly),     
two Higgs doublets $\varphi_{1,2}$, and in addition 
an extra superfield  $\Psi$ to be identified with 
the right-handed neutrino, which is a gauge singlet 
of the standard model. For simplicity, let us 
consider only one fermion generation.

The charged lepton gets mass from the Yukawa term $hl e^c \varphi_1$, 
while the similar coupling $gl\Psi\varphi_2$ induces the Dirac mass 
for the neutrino component. For implementing the hybrid inflation 
scenario, we consider the superpotential of the following form 
\cite{zurab}:
\begin{equation} \label{newsuper}
W=-\Lambda^2S+\lambda S \Phi^2+\kappa \Phi\Psi^2\,, 
\end{equation}
which is the simplest modification of the original one ~(\ref{super}). 
It is essential the term $\kappa\Phi\Psi^2$ which 
communicates the standard particle sector to the fields 
producing inflation. 
The above superpotential as well as the Yukawa terms respect the 
global $R$-symmetry with the fermion superfields 
$\Psi,l,e^c,...$ carrying the $R$ charge $1/2$ and the 
Higgs ones $S$ and $(\Phi,\varphi_{1,2})$ carrying charges 
$1$ and $0$ respectively. Notice that this $R$ symmetry 
forbids the R-violating terms $l\varphi_2$, $lle^c$ etc. - in other words, 
the theory has an automatic matter parity under which the 
all fermion superfields change sign while the Higgs ones remain invariant.
Interestingly, $R$ symmetry forbids also the supersymmetric mass  
term $\mu\varphi_1\varphi_2$ which seems to be a good starting 
point for solving the hierarchy problem. Needless to say, 
we assume that the Higgsino mass can emerge in some ways
in consequence of the supersymmetry breaking.
Moreover, in global supersymmetry limit, the only mass scale in the
theory is $\Lambda$ in~(\ref{newsuper}).

Let us take the scalar field components in a form: 
\begin{equation} \label{scalars}
S=\frac{\sigma}{\sqrt{2}}\,, ~~~~
\Phi=\frac{\phi}{\sqrt{2}}\,, ~~~~
\tilde\Psi =\frac{\psi_1 + i\psi_2}{\sqrt{2}}\,, 
\end{equation}
where $R$-symmetry transformation has been used to make $S$ field real, 
and the imaginary part of $\Phi$ has been put to zero for simplicity. 
(Sometimes it is convenient to present the last field in 
terms of its modulus and phase: 
$\tilde\Psi=(\rho/\sqrt2) \exp(i\delta)$.) 
Their potential reads: 
\begin{eqnarray}
\label{pot1} 
V&=& V(\sigma,\phi) +\kappa^2\phi^2(\psi_1^2+\psi_2^2)    
+\frac{\kappa^2}{4}(\psi_1^2+\psi_2^2)^2 
\nonumber \\
& & + \kappa\lambda\sigma\phi(\psi_1^2-\psi_2^2) \, ,  
\end{eqnarray} 
where $V(\sigma,\phi)$ is given by~(\ref{pot0}). 
It has a global minimum with a non-zero VEV $\langle\Phi\rangle$: 
\begin{equation} \label{min} 
\langle \sigma \rangle = 0, ~~~ 
\langle \psi_{1,2} \rangle = 0, ~~~ 
\langle \phi \rangle = \phi_0 = \sqrt{2}\lambda^{-1/2}\Lambda , 
\end{equation}
At this minimum the superfields $S$ and 
$\Phi^\prime = \Phi -\langle \Phi\rangle$ have equal masses 
\begin{equation} \label{Mphi} 
M_\sigma = M_\phi = 2 \lambda^{1/2} \Lambda \simeq 
\eta \times 2.5 \cdot 10^{15} ~{\rm GeV}\,,  
\end{equation} 
while the last coupling in (\ref{newsuper}) induces the mass 
of $\Psi$: 
\begin{equation} \label{Mpsi}
M_\Psi  = \frac{2\kappa \Lambda}{\lambda^{1/2}} \simeq 
\frac{\kappa\eta }{\lambda} \times 2.5\cdot 10^{15} ~{\rm GeV}\,, 
\end{equation}
where  (\ref{eta}) has been used for numerical estimations.  
Interestingly, this is just the right range for the RH  
neutrino mass. Then, by means of the seesaw mechanism,   
the ordinary neutrino gets small Majorana mass  
\begin{equation}
\label{numass}
m_{\nu} = \frac{g^2 \langle\varphi_2\rangle^2}{ M_{\Psi}} 
\simeq \frac{g^2\lambda}{\kappa \eta} \times  
1.2 \cdot 10^{-2} ~{\rm eV}\,,
\end{equation} 
where for numerical estimation we have taken 
$\langle\varphi_2\rangle \simeq 170$ GeV 
(i.e. not very small $\tan\beta$). 

The superpotential~(\ref{newsuper}) provides the same  
pattern for the inflation as the original one~(\ref{super}). 
For the field values $\sigma > \sigma_c = \lambda^{-1/2}\Lambda$, 
the potential~(\ref{pot1}) has a flat direction along the 
$\sigma$ axis, in which the orthogonal field is trapped in 
false vacuum with the energy $V(\sigma, 0) = \Lambda^4$.    
This flat direction is lifted by radiative corrections resulting 
from the supersymmetry breaking by the non-zero 
vacuum energy, and perhaps also by some other supergravity 
corrections \cite{susy1}. 
The precise form of these corrections is not important, 
and we can simply assume that they result in an effective 
$\sigma$ dependent potential, e.g. in the form of 
mass term, $\sim \bar{m}^2\sigma^2$, $m \ll \Lambda$,     
that gives an appropriate slope to the inflaton 
potential on which $\sigma$ can slowly roll down and 
produce inflation. For achieving this, the parameters 
\begin{eqnarray} \label{slow} 
&&
\epsilon = \frac{M_{\rm P}^2}{16\pi}
\left(\frac{V^\prime}{V}\right)^2 \simeq 
\frac{\lambda^2 M_{\rm P}^2\sigma^2}{\pi \Lambda^4}
\left(\frac{\bar m}{M_\sigma}\right)^4 \,, 
\nonumber \\ 
&&
\eta = \frac{M_{\rm P}^2}{8\pi}
\left(\frac{V^{\prime\prime}}{V}\right) \simeq 
\frac{\lambda M_{\rm P}^2}{2\pi \Lambda^2}
\left(\frac{\bar m}{M_\sigma}\right)^2 \,, 
\end{eqnarray} 
have to be much smaller than one. For producing inflation, 
the orthogonal field should be settled in a false 
vacuum $\phi=0$ while the inflaton evolves from an 
initial value $\sigma \sim M_{\rm P}$.
The inflationary phase ends up when the inflaton approaches 
the critical value $\sigma_c$, after which both fields 
$\sigma$ and $\phi$ begin to oscillate near their global minima 
and reheat the universe.

Let us now turn to the role of the superfield $\Psi$. 
In the global minimum~(\ref{min}), 
its fermion component, the RH neutrino, gets large 
Majorana mass and thus violates the lepton number conservation. 
However, the Lagrangian of the scalar component $\tilde{\Psi}$, 
the RH sneutrino, maintains the global $U(1)$ symmetry.
Indeed, since $\sigma=0$, the scalar potential~(\ref{pot1}) 
becomes a function of the field $\tilde\Psi$ modulus 
$\rho = (\psi_1^1+\psi_2^2)^{1/2}$ and does not depend 
on its phase $\delta$. 
Needless to say, that the Yukawa coupling to lepton and 
Higgsino, $\tilde{\Psi}l\tilde\varphi_2  + {\rm h.c.}$,  
are conserving the lepton number. 

On the other hand, in the false minimum $\phi=0$,  
the superfield $\Psi$ is massless and 
completely decoupled form the inflaton $\sigma$. 
Now the scalar potential of $\tilde\Psi$
contains only the last term $\propto \kappa^2\rho^4$ 
in~(\ref{pot1}).   
Thus, the global $U(1)$ symmetry associated with the lepton 
number is restored. 

Therefore, the $U(1)$ symmetry in the potential of 
RH sneutrino fields $\tilde\Psi$ 
can be violated only by the last term in~(\ref{pot1}),  
when that both $\phi$ and $\sigma$ have non-zero values. 
This can occur only during the epoch when the background 
fields $\phi,\sigma$ oscillate near their global minima. 
As we show below, at this phase the non-zero lepton number 
can be generated in the classical motion of the RH sneutrino 
fields, which will be transferred to the standard 
leptons via their decay $\tilde\Psi\to l\tilde\varphi_2$.

%%%%%%%%%%%%%%%%%%%%%%%%%%%%%%%%%%%%%%%%%%%%%%%%%%%%%%%%%%%%%%%%%%%%%%%%%%

\section{Dynamics of the fields }

In this section we shall describe the evolution of the fields 
in our model. Starting our consideration 
from an earliest time when the Planck era ends 
and classical general relativity becomes applicable, 
we assume that at this moment all classical fields 
have initial values $\sim M_{\rm P}$  
and hence the energy density of the Universe is 
$\sim M_{\rm P}^4$, quite a generic situation. 
We study the scalar field dynamics before the inflation onset, 
during exponential expansion, their postinfaltionary oscillations 
and, finally, we  discuss the reheating of the Universe.   
This will provide us all necessary tools 
to calculate the lepton asymmetry. 

\subsection{Pre-inflation}

The detailed analysis of the scalar field dynamics before 
inflation in the model with the superpotential 
~(\ref{newsuper}) has been performed in Ref.~\cite{zurab}. 
Here we briefly recall its main features. 

All scalar fields $\sigma$, $\phi$ and $\psi_{1,2}$ 
have initial values of close to the Planck scale, and 
with the Hubble expansion they evolve down due to interaction 
terms in~(\ref{pot1}). Therefore, the cosmological energy 
density, initially $\sim M_{\rm P}^4$, 
decreases with the Hubble expansion and  its evolution settles 
quickly in a pattern when the energy of the system is dominated 
by the regular oscillations of $\phi$ around zero. 
From this moment the Universe expands as in a matter dominated era. 

Along the flat direction the effective mass for $\phi$ field 
is quite large,  
$M_{\phi,{\rm eff}}^2 = 2\lambda^2(\sigma^2-\sigma_c^2)$, 
and it vanishes only when $\sigma = \sigma_{c}$. 
On the other hand, once $\phi$ has a large amplitude, 
it induces the large curvature for the inflaton field $\sigma$ 
and the latter starts to quickly roll down to the origin. 
Therefore, in order to produce inflation, 
the amplitude of $\phi$ must settle to zero before $\sigma$ 
will manage to substantially decrease its initial 
value $\sim M_{\rm P}$. 
The friction provided by the Universe expansion itself 
cannot really help, since the Hubble parameter 
in this epoch is actually smaller than $M_{\phi,{\rm eff}}$  
and thus the field $\phi$ suffers 
many oscillations during the Hubble time.

However, the problem can be with help of the coupling 
$\kappa\Phi\Psi^2$, which permits the oscillating field 
$\phi$ to decay into $\Psi$ particles. The decay rate 
$\Gamma(\phi\to\Psi\Psi) =(\kappa^2/8\pi)M_{\phi,{\rm eff}}$  
can be large as far as the coupling constant is reasonably 
large, $\kappa > 0.1$ or so, and $\sigma \sim M_{\rm P}$. 
In addition, as soon as $\phi$ becomes zero, $\sigma$  
stops to feel the classical fields $\psi_{1,2}$ as 
well as their quanta produced in the decay,  
large curvature of the inflaton potential  disappears and 
its value freezes at values $\sigma \sim M_{\rm P}$. 
By the time $t\sim 1/\Gamma$ the cosmological energy density 
becomes dominated by the relativistic $\Psi$ particles, 
and it fastly decreases with the universe expansion 
until it becomes dominated by the false vacuum energy. 
Since this moment Universe starts to expand exponentially 
while the inflaton proceeds its slow roll due to 
small curvature term.

The decay of $\phi\to 2\Psi$ can be understood as a process 
of preinflationary reheating after which the system from 
a generic initial state, with all fields having 
magnitudes $\sim M_{\rm P}$, settles in false vacuum,  
Therefore, it provides a simple mechanism to obtain the
correct initial condition for the inflation to take place. 
Once inflation commences it dilutes the produced 
$\Psi$ quanta.  

Let us turn now to the evolution of the classical 
sneutrino field $\tilde\Psi$~(\ref{scalars}).
As far as we did not introduce any mass term for the RH neutrino, 
in the superpotential~(\ref{newsuper}), 
the fields $\psi_{1,2}$ are intrinsically massless. 
Nevertheless, before the decay of $\phi$ oscillations,  
they have field dependent effective mass terms
$m^2_{1,2}=\kappa^2\phi^2 \pm \kappa\lambda \sigma\phi$  
and therefore during the matter dominated phase 
prior to inflationary phase they simply roll down from the 
initial values order $M_{\rm P}$ .  
However, when $\phi$ settles to zero and inflation begins, 
these fields become massless and their oscillations are damped. 
By this moment these fields still have reasonably big 
values. However, they have further non-trivial evolution 
during and after inflation, which we shall discuss in subsequent 
sections. 

%%%%%%%%%%%%%%%%%%%%%%%%%%%%%%%%%%%%%%%%%%%%%%%%%%%%%%%%%%%%%%%%%%%%

\subsection{Inflation}

As was told in the above, the inflation proceeds when 
$\phi$ is set to zero while $\sigma$ still has a large 
value $\sim M_{\rm P}$. 
On the other side, $\sigma$ field has a small effective
mass term $\sim \bar{m}^2\sigma^2$ which allows it to roll down 
the potential. The main contribution to the energy
density comes from the false vacuum, 
$V(\sigma,0)\simeq \Lambda^4$, 
and the Universe expands exponentially up to the time $t$ when 
the inflaton field reaches the critical value  $\sigma_{c}$, 
where on the Universe exits gracefully from almost de-Sitter 
expansion during which the cosmological scale has grown 
by a factor $\exp(Ht)$.  
The Hubble parameter during the inflation is given by  
\begin{equation}\label{H}
H \simeq 
\sqrt{\frac{8\pi}{3}}\frac{\Lambda^2}{M_{\rm P}} \simeq 
\frac{\eta^2}{\lambda}\times 4 \cdot 10^{11} ~{\rm GeV}.  
\end{equation} 
As soon as  $\phi$ settles to zero and inflation begins, 
$\psi_{1,2}$ become massless and the curvature of their 
potential is induced only by quartic self-couplings in~(\ref{pot1}).  
These fields are completely decoupled from the inflaton 
and thus they evolve almost independently during the inflationary 
era, with the  following equations of motion: 
\begin{equation}
\label{dyn1}
\ddot \psi_{1,2}+3H\dot\psi_{1,2} = 
-\kappa^2(\psi_1^2+\psi_2^2)\psi_{1,2} \,, 
\end{equation}
or, in terms of the modulus $\rho$, 
$\ddot \rho +3H\dot\rho = -\kappa^2\rho^3$. 
As for the phase $\delta$, it essentially becomes a flat 
mode as far as $\phi=0$ and the lepton number conservation 
is restored.

The classical field $\rho$ keeps rolling down  
until it approaches values $\sim H$. 
After this  its dynamics is almost frozen and 
the rest of  evolution is determined by the equation 
$3H\dot \rho \approx -\kappa^2 \rho^3$.  
From here one immediately obtains that by the time $t$   
when the slow roll ends up, the field value will be 
$\rho \approx \kappa^{-1}(3H/t)^{1/2}$. This can be 
rewritten as 
\begin{equation} \label{val}
\rho = \frac{\sqrt{C}}{\kappa} H \, , 
\end{equation}
where $C= 3/N_{\rm e}$, with  $N_{\rm e}=H t$ being 
the total number of e-foldings. 
Notice also, that in hybrid inflation models, 
unlike chaotic models, the number of e-foldings can not be 
arbitrarily large and for our purpose we assume that maximum 
number of e-folding can at most be $N_{\rm e} \approx 100$.

So far in the evolution of $\Psi$ fields we have not taken into
consideration any kind of supergravity correction to their masses. 
In particular, for a generic K\"ahler potential 
the fields $\tilde\Psi$ could get the mass term 
$- C H^2 (\psi_1^2+\psi_2^2)$, $C$ being order 1 coefficient.   
The sign of $C$ is model dependent and it 
can not be determined correctly. 
If the correction is positive, then during inflation 
$\rho$ evolves as $\propto e^{-3Ht}$ and it 
will essentially vanish at the inflation exit.  
However, if the mass correction turns out to be negative,
then $\tilde\Psi$ fields will have a false 
minimum with a non-zero VEV which breaks the lepton number:  
$\langle\rho\rangle^2=CH^2/\kappa^2$.  
Therefore, during inflation these fields will fastly evolve down 
until being trapped in the false minimum, 
and remain stuck to it until the effective mass of 
$\Psi$ field overtakes the expansion parameter $H$. 
This happens very soon after the end of inflation, 
because $M_\Psi^2\propto \Lambda^2 \gg H^2$. 
Therefore, by the end of the inflation era
$\rho$ has a non-vanishing magnitude which can be given 
still by~(\ref{val}), but this time $C$ being some unknown 
order 1 coefficient.

%%%%%%%%%%%%%%%%%%%%%%%%%%%%%%%%%%%%%%%%%%%%%%%%%%%%%%%%%%%%%%%%%%%%%%%
\subsection{Post-inflation}

The evolution of $\sigma$ and $\phi$ after the inflation exit 
has been studied in Refs.~\cite{mar1,dan}. 
After the phase transition the fields oscillate with more or less 
similar amplitude near their global minima~(\ref{min}), 
the maximum amplitude attained by $\sigma$ field is
$\sigma_{c}=\phi_0/\sqrt{2}$, while $\phi$ takes at most 
$\phi_0$. The initial conditions for the oscillations are fixed by 
the inflationary dynamics and are given by~\cite{mar1,dan}: 
\begin{eqnarray}\label{i}
\sigma_{\rm i} &=& \sigma_c \pm \frac{H_{\rm i}}{2\pi}\,, ~~~~~~
\phi_{\rm i}=\frac{H_{\rm i}}{2\pi}\,, 
\nonumber \\ 
\dot \sigma_{\rm i} &=& -\frac{1}{3H_{\rm i}}
\frac{\partial V}{\partial \sigma}\,, ~~~~~
\dot \phi_{\rm i} = -\frac{1}{3H_{\rm i}}
\frac{\partial V}{\partial \phi}\,,
\end{eqnarray}
where the initial velocities are calculated at $\phi_{\rm i}$ and 
$\sigma_{\rm i}$, and $H_{\rm i}$ is the Hubble  parameter 
at the end of inflation.

The oscillations are essentially an-harmonic in nature. However, 
near the global minimum, there exists a particular solution which
satisfies a straight line trajectory in $\phi-\sigma$ plane 
\cite{mar1,dan};
\begin{equation}
\label{an1}
\phi = \sqrt{2}(\sigma_{c} - \sigma)\,.
\end{equation}
Near the bottom of the potential the oscillations are harmonic and 
their frequency is governed by the mass 
of the fields at their minima, see~(\ref{Mphi}). 
It is possible to give an approximate analytical solution
\begin{equation}
\label{an2}
\frac{\phi(t)}{\phi_0}\approx 1+\frac{A(t)}{3}\cos(m_{\phi}t)\,,
\end{equation}
where $A(t)\sim 1/t$ is slowly time-varying amplitude of the oscillations,
which depends  crucially upon the ratio $H_{\rm i}/M_\phi$.
The smaller the ratio is, the larger is the number of oscillations of 
the fields before they feel the Hubble expansion. 
For smaller inflationary scales, such as $\Lambda \leq 10^{-3} M_{\rm P}$,  
the ratio $H_{\rm i}/M_\phi$ is small, 
$H_{\rm i}/M_\phi\simeq 1.3\Lambda/\lambda^{1/2}M_{\rm P}\ll 1$.  
This is an interesting characteristics of the supersymmetric hybrid
inflation model, which tells  us that there are many oscillations 
of the background fields with an almost constant amplitude. 
The effect of expansion is felt after many oscillations, and this has 
been verified numerically in Refs.~\cite{mar1,dan,mar2}.

Let us consider now post-inflationary dynamics of the 
classical fields $\tilde\Psi$. 
Although they have been essentially the massless fields during 
the inflation, after inflation they become massive, with mass 
$M_\Psi \sim \Lambda$. In other words, 
$\psi_{1,2}$  fields get quadratic terms in the potential 
and they start to oscillate near the origin with the 
frequencies governed by mass $M_\Psi$~(\ref{Mpsi}). 
Depending on the situation, 
the initial amplitudes for these oscillations are the field 
which they have at the exit of inflation, see~(\ref{val}).  
For keeping more generality, 
let us parameterize the magnitudes of these amplitudes as 
\begin{equation}
\label{in}
\psi_{1(2)}^{\rm i} = \rho_{\rm i} \cos\delta_{\rm i} 
(\sin\delta_{\rm i})\,, 
~~~~ \rho_{\rm i}^2 = \frac{C}{\kappa^2} H  
\end{equation}
where we assume that the phase $\delta$ is order one, 
and $C = 3/N_{\rm e}$ or $C\sim 1$, depending on the situation 
whether at inflation stage these fields had  
the supergravity induced order $H^2$ mass terms or not. 
Therefore, the classical fields $\psi_{1,2}$ have 
the initial amplitudes $\sim H$,  
while $\sigma$ and $\phi$ fields have much larger amplitudes 
$\sim \Lambda/\sqrt{\lambda}$. 
The initial energy density $V\simeq \Lambda^4$ is 
dominated by oscillations of $\sigma$ and $\phi$, 
while the contribution of $\Psi$ is negligible. 
In this way, one can safely neglect backreaction 
of $\tilde\Psi$ fields on the oscillation  
of the classical background fields $\sigma$ and $\phi$.

The post-inflationary evolution of classical $\tilde\Psi$ 
fields is quite interesting. The oscillation frequency of these fields 
is governed by the value of $\phi$, 
and also the evolution of $\sigma$ leads to an additional 
contribution to their equation motion. On the other hand, 
once they become massive and can decay into 
leptons and higgsinos, the decay rate~(\ref{Gamma}) 
also contributes the friction term in their equation of motion,       
 which now read 
\begin{eqnarray}
\label{dyn3}
\ddot \psi_{1,2} +(3H+\Gamma) \dot\psi_{1,2} = 
& -\kappa^2(2\phi^2+\psi_1^2+\psi_2^2)\psi_{1,2} 
\nonumber \\
& \mp 2\kappa\lambda\sigma\phi\psi_{1,2} \,,   
\end{eqnarray}
Notice, that the last terms in the above equations come 
with an opposite sign and so the evolution of $\psi_1$ 
is different from $\psi_2$. 
In other words, in the background of the $\sigma,\phi$ fields,  
$\psi_{1,2}$ get the 
$U(1)$ invariant effective mass term  
$M_+^2(\phi)\tilde\Psi^\ast\tilde\Psi =
2\kappa^2\phi^2 (\psi_1^2 + \psi_2^2)$, 
as well as the $U(1)$ violating one 
$M_{-}^2(\sigma\phi)(\tilde\Psi^{\ast 2} + \tilde\Psi^2) = 
2\lambda\kappa\sigma\phi(\psi_1^2 -\psi_2^2)$.  

As we have already remarked,  $\phi,\sigma$ fields make many 
oscillations in one Hubble time, which  
allows us to consider the average effect of these 
fields upon $\psi_1,\psi_2$ fields. In other words, 
one can replace $\phi^2$ and $\sigma\phi$ by their mean 
values within a period of one Hubble time,  
for which from~(\ref{an1}) and~(\ref{an2}) we obtain:  
\begin{equation} \label{average} 
\langle\phi^2\rangle_t \simeq \frac{2\Lambda^2}{\lambda} , 
~~~~
\langle \sigma\phi\rangle_t \simeq 
\frac{\sqrt2 \Lambda^2}{18\lambda} A^2(t) ,  
\end{equation} 
where $A(t)^2 \propto 1/t^2$.
Substituting these averages in the~(\ref{dyn3}), we see that during posinflationary 
oscillations $\psi_1$ and $\psi_2$ have 
different dynamical mass terms: 
\begin{equation} \label{M12}
M^2_{1,2}(t) = M_+^2 \pm M_-^2 = M_\Psi^2 
\left(1 \pm \frac{A(t)^2}{25}\right)  
\end{equation} 
This dynamical mass splitting causes their helical motion 
in the background of $\sigma$ and $\phi$  and produces the 
lepton asymmetry in our model.

The evolution $\tilde\Psi$ fields crucially depends 
whether the friction term in~(\ref{dyn3}) is
dominated by $H$ or $\Gamma$. 
In the former case, the amplitude 
$\psi$ decreases with time as $\propto 1/t$, while in the latter 
case as $\exp(-\Gamma t/2)$. For the width of the decay 
$\tilde\Psi\to l\tilde\varphi_2 (\tilde{l}\varphi_2)$ we have: 
\begin{equation}\label{Gamma}  
\Gamma_\Psi = \frac{g^2}{8\pi} M_\Psi \simeq 
\left(\frac{\kappa\eta}{\lambda}\right)^2
\left(\frac{m_\nu}{0.1~ {\rm eV}}\right) 
\times 10^{15}~{\rm GeV}. 
\end{equation}

%%%%%%%%%%%%%%%%%%%%%%%%%%%%%%%%%%%%%%%%%%%%%%%%%%%%%%%%%%%%%%%%%%%%%%%%%%%%%%%
\subsection{Reheating}

The end of inflation also marks an era of entropy creation in the 
Universe. The energy stored in oscillations of classical fields 
$\phi$ and $\sigma$ eventually decays into relativistic particle species. 
In our case reheating could occur via two possible channels.
Either $\sigma$ and/or $\phi$ have some dominant channel to 
decay into standard particles, or, $\phi$ could decay into 
right-handed neutrino $\Psi$ quanta, which would subsequently 
decay into leptons and Higgs to produce a final thermal bath.
The choice of the dominant channel shall depend on the 
couplings in the theory which in our case are constrained 
to guarantee that the reheat temperature of the Universe 
is not be too large. In particular, in supersymmetric theories 
there is an upper bound on the reheating temperature 
$T_{\rm r} \leq 10^{9}$ GeV or so \cite{sarkar}. 
If this condition is not met, then the gravitinos are effectively 
produced via scattering processes in thermal bath, and their 
late decay products can genuinely threaten the standard 
nucleosynthesis unless there is some other mechanism 
to dilute their number density. 

The reheating temperature of the Universe era can be estimated as 
\begin{equation}
T_{\rm r} \approx 0.5 g_\ast^{-1/4} 
\sqrt{\Gamma  M_{\rm P}} \simeq 0.1 \sqrt{\Gamma  M_{\rm P}}\,,
\end{equation}
where $\Gamma$ is the decay rates of 
the fields $\sigma$ or $\phi$, and $g_\ast$ is an effective number of 
the particle degrees of freedom in the thermal bath. 
(In the supersymmetric standard model which we consider here, 
$g_\ast \sim 200$).

The following remark is in order. 
Apart from the perturbative decay of the oscillating fields, 
the coherent oscillations of the inflaton could also lead to 
a non-thermal resonant production of particles \cite{kofman}, 
and in particular of gravitinos with helicity $3/2$~\cite{maroto},
and helicity $1/2$~\cite{grav}. This would produce very stringent 
upper bound on the reheating temperature. 
However, it has been realised later on that non-thermal 
production of gravitinos during inflation occurs via 
the goldstino mode, a helicity $\pm 1/2$ component which can 
be recognised as inflatino. It eventually 
decays along with the inflaton to reheat the Universe, 
and thus non-perturbative production of gravitinos should not 
pose serious problems for nucleosynthesis \cite{ruz}.

One of the most important criteria for a resonant particle 
production is the coherent oscillations of the background fields,
which is fulfilled in supersymmetric hybrid inflationary model.
It has been noticed that the particle creation is quite
efficient in supersymmetric hybrid models, see Ref.~\cite{mar1}. 
In particular, quanta of the heavy RH neutrinos produced 
at the preheating stage by non-perturbative decay of inflaton 
oscillations, in their subsequent decays could produce the 
lepton asymmetry, provided that these decays are CP-violating 
\cite{giudice}.

Another interesting feature of hybrid model which we must mention 
here is the possibility to realise tachyonic preheating. This is  
because near the critical point mass squared for $\phi$ field 
flips its sign and becomes a tachyonic mode. 
This violates the adiabatic vacuum condition and leads to
explosive production of particles \cite{felder}. 
In such a case, it is also important to consider the 
backreaction of the quanta on the classical fields, which 
actually leads to destabilising the system and the zero-mode 
trajectories \cite{dan}.
Eventhough, tachyonic preheating might work, it is still not 
very clear why there should be a resonant particle production 
once the backreaction of the quanta is properly taken into account.

Whatsoever be the case,  we shall not worry too much 
upon the preheating aspects. 
All we assume here is that the decay of $\phi$ and $\sigma$ is
responsible for reheating the Universe. 
We also neglect the finite temperature effects, which 
could emerge if the relativistic particle species produced 
by the inflaton decay thermalize too early, before the 
inflaton decay ends up, and thus can give thermal corrections 
to the potential of the fields such as $\tilde\Psi$ in our case.
The finite temperature mass corrections can be avoided once 
thermalization of the Universe is delayed until 
the last stages of reheating.

Coming back to our situation, we have to control that the 
already existing couplings in the model will not produce too 
large reheat temperature. The dominant channel for the 
decay of $\Phi$ field  is governed by the last term in the 
superpotential~(\ref{newsuper}). If $M_\phi > M_\Psi$, 
then $\Phi$ decays into $\Psi$ quanta which then produce 
the standard particles via the coupling 
$g\Psi l \varphi_2$. In this case we have 
$\Gamma_\phi= (\kappa^2/8\pi) M_\phi$ and thus 
$T_{\rm r} \sim 0.1 \sqrt{\Gamma_{\phi} M_{\rm P}} 
\simeq 3\kappa\eta^{1/2} \times 10^{15}$ GeV, 
where we have used~(\ref{Mphi}). Taking now into account the 
fact that in our approach the constant $\kappa$ should 
be rather large , since otherwise the decay 
of the orthogonal field $\phi$ would not be effective in 
the preinflationary phase, and neither $\eta$ can be 
be very small, we obtain too a big reheat temperature.

However, if $M_\Psi > M_\phi$, i.e. $\kappa > \lambda$. 
situation is more interesting and the problem can be solved 
without any extra assumptions. 
Now the decay channel $\Phi\rightarrow \Psi\Psi$  
is kinematically forbidden, and the lowest order 
relevant operator for the decay of $\phi$ into lighter particles 
is the $D=6$ one $(\kappa g^2/M_\Psi^2)\Phi l^2 \varphi_2^2$. 
Such term in the superpotential is induced by the 
Yukawa couplings $g\Psi l \varphi_2$  
after integrating out the heavy superfields $\Psi$. 
Now the decay width of the 
four-body decay $\Gamma(\phi \rightarrow ll\varphi\varphi) 
= P \kappa^2 g^4 M_\phi^5/M_\Psi^4$ is suppressed 
by a phase space factor $P\sim 10^{-5}$. Finally, by 
taking into account~(\ref{Mphi}), 
(\ref{Mpsi}) and (\ref{numass}), 
the reheating temperature comes naturally as
\begin{equation}\label{tr}
T_{\rm r} \simeq \lambda \eta^{3/2} 
\left({m_\nu\over 0.1~{\rm eV}}\right) \times 10^{14}~\mbox{GeV} .  
\end{equation}
For instance, by taking both $\eta$ and $\lambda$ 
order $10^{-2}$ and $m_\nu\sim 0.1$ eV,  
we get $T_{\rm r}\sim 10^9$ GeV.  
This result also satisfies  the thermal gravitino production 
bound on the reheating temperature. 
In addition, as we shall see in the next section, 
such a reheat temperature can easily produce 
the right amount of the baryon asymmetry.

%%%%%%%%%%%%%%%%%%%%%%%%%%%%%%%%%%%%%%%%%%%%%%%%%%%%%%%%%%%%%%%%%%%%%%%%

\section{Lepton asymmetry}

As we have mentioned earlier, the RH sneutrino field 
$\tilde\Psi$ carries the global $U(1)$ charge, 
and the associated quantum number $B-L$ (rather than $L$) 
is violated via the last term  in the potential (\ref{pot1}).
This term is effective during the post-inflationary oscillations 
and it gives rise to a helical motion of $\psi_{1,2}$ 
in the background of $\sigma$ and $\phi$ fields, 
thereby generating the lepton number 
in the postinflationary universe.

The $B-L$ charge density generated during the time evolution 
of the RH sneutrino field is a zero component of the 
global $U(1)$ current: 
\begin{equation} \label{current}
n_{\rm B-L}= 
\frac{i}{2}({\dot{\tilde\Psi}}^\ast \tilde\Psi-
\tilde\Psi^\ast\dot{\tilde\Psi}) = 
\frac12(\psi_1\dot\psi_2 - \dot\psi_1 \psi_2 ) \,. 
\end{equation}
Then from  (\ref{dyn3}) we immediately obtain the
equation which describes its evolution: 
\begin{equation}
\label{lep}
\dot n_{\rm B- L}+3Hn_{\rm B-L}=2\kappa\lambda
\sigma(t)\phi(t)\psi_1(t)\psi_2(t) \,.
\end{equation}
where the RH side acts as a source term which
generates a net $B-L$ asymmetry through a non-trivial 
motion of $\psi_1$ and $\psi_2$ fields. 

Let us integrate this equation from the time moment $t_{\rm i}$, 
which corresponds to the end of inflation, 
up to a finite time interval $t$: 
\begin{equation}
\label{integral}
n_{\rm B-L} = \frac{2\kappa \lambda}{R_t^3}  
\int_{t_{\rm i}}^{t} dt^\prime R^3_{t^{\prime}} 
\langle\sigma\phi\rangle_{t^{\prime}} 
\rho^2(t^{\prime})\sin2\delta_{t^\prime} 
\end{equation}
where $R_t$ is scale factor and we have substituted the mean 
value $\langle\sigma\phi\rangle_t$ given 
in~(\ref{average}). The CP-phase $\delta_t$ changes 
slowly, since for $\lambda<\kappa$ oscillations of the 
fields $\psi_1$ and $\psi_2$ have about the same 
oscillation frequency (see~(\ref{M12}). 
In terms of the field quanta, $n_{B-L}$ is nothing but
a number density difference between the RH sneutrino 
$\tilde\Psi$ and anti-sneutrino $\tilde\Psi^\ast$ states.
It will be transmitted to the standard particle system 
via the decay of the RH sneutrinos into the 
ordinary leptons and Higgsinos (or sleptons and Higgses), 
Therefore, even if the decay rates
$\tilde\Psi \to l \tilde\varphi_2$ 
and $\tilde\Psi^\ast \to \bar{l} \tilde\varphi_2$, 
are exactly the same (no CP-violation in decays), 
we produce the different amount of $l$ and $\bar{l}$.

Once the $B-L$ is non-zero, the net baryon number is induced 
via sphaleron effects which 
violate $B+L$ but conserve $B-L$ \cite{thooft}.  
The sphalerons are active in a temperature range from about 
$10^{12}$ GeV down to 100 GeV. In the context of the 
supersymmetric model the relation between the $B$ and $B-L$ 
is given by $B = - 0.35 (B-L)$ \cite{kleb}.  
Therefore, for obtaining the observed baryon to entropy density ratio 
in the range  $B=n_B/s = (0.3 - 1)\cdot 10^{-10}$, 
as it is restricted by the primordial nucleosynthesis bounds, we 
need $B-L \sim {\cal O}(10^{-10})$.

Now we are in grade to calculate the $B-L$ number to 
entropy density ratio $B-L = n_{B-L}/s$ in the Universe. 
Assuming that the entropy is generated at the postinflationary 
reheating and there is no more entropy injection at later times, 
we calculate from~(\ref{integral}) the value of $n_{B-L}$ 
produced by the reheating time 
$t=t_{\rm r}\simeq 0.3 g_\ast^{-1/2} M_{\rm P}/T_{\rm r}^2$,  
and compare it to the entropy density at this time,  
$s=(2\pi^2/45)g_{\ast}T_{\rm r}^3$.

The Universe dominated by the field oscillations 
expands as in a matter dominated era, 
and so the scale factor changes as $R_t\propto t^{2/3}$.  
On the other hand, $\langle\sigma\phi\rangle_t \propto 1/t^2$  
while the CP-phase $\delta_t$ changes 
slowly, since for $\lambda<\kappa$ oscillations of the 
fields $\psi_1$ and $\psi_2$ have almost the same oscillation 
frequency. Therefore, the integrand in~(\ref{integral}) goes as 
$\rho^2(t)$, and the final answer depends 
whether the decay rate~(\ref{Gamma}) is larger or smaller 
as compared to the Hubble parameter. Let us consider first the case $\Gamma < H$.  
As we discussed in previous section, 
then we have $\rho^2 \propto 1/t^2$. 
This suggests that the maximum contribution to integral~(\ref{integral}) 
comes at the initial times $\sim 1/H_{\rm i}$ and hence we obtain:
\begin{equation} \label{B-L}
B-L  \simeq 
\frac{\kappa \rho^2_{\rm i}}{H^2} \frac{T_{\rm r}}{M_{\rm P}} 
\sin2\delta_{\rm eff}\,,
\end{equation}
where $\delta_{\rm eff}\sim 1$ is an effective $CP$ violating 
phase which we assume to be ${\cal O}(1)$, and  
$\rho_{\rm i}$ is given by~(\ref{in}).  

However, if the friction term in~(\ref{dyn3}) is 
dominated by the decay width, i.e. $\Gamma > 3H$, 
then the RH sneutrino field behaves like 
$\rho^2 \propto \exp(-\Gamma t)$ and,  
as a result, the magnitude of $B-L$ will be 
reduced by a factor $x = 3H/\Gamma$   
which can be obtained from~(\ref{H}) and (\ref{Gamma}): 
Therefore, the final result for the lepton
asymmetry reads: 
\begin{equation} \label{final} 
B-L \sim X \frac{T_{\rm r}}{M_{\rm P}} \simeq X 
\left(\frac{T_{\rm r}} {10^{9} ~{\rm GeV}}\right) \times 10^{-10} \,, 
\end{equation} 
where the numerical factor is 
$X = \min\{1, x\}C\kappa^{-1}$, with $C$ being $\sim 1$ 
coefficient if the field $\tilde\Psi$ gets supergravity 
induced mass term during inflation, or 
$C = 3/N_{\rm e}$ otherwise.  

For $X\sim 1-10$, the result~(\ref{final}) implies 
correct magnitude of the baryon asymmetry if 
$T_{\rm r}\sim 10^9$ GeV or so. This range for the reheating 
temperature can be natural in the context of our model, 
provided that the factor $\lambda\eta^{3/2}$ in  (ref{tr}) 
is ${\cal O}(10^{-5})$. 
The larger $T_{\rm r}$ would contradict 
to the non-thermal gravitino production limit.
On the other hand, if $x > 1$, then the coefficient $X$ order 
1 or 10 can easily emerge if $\kappa \sim 10^{-1}$ 
and $C \sim 1$ or $1/20$, the latter value attained 
to the case of 60 e-fold inflation.

%%%%%%%%%%%%%%%%%%%%%%%%%%%%%%%%%%%%%%%%%%%%%%%%%%%%%%%%%%%%%%%%%%%%%%%%%%%%%%%%%%%%%%
\section{Discussion}

Up to now we have considered only one fermion generation. 
Let us incorporate now all three generations and discuss 
what happens in this case. In other words, we introduce 3 
lepton species 
$l_a, e^c_a$ and $\Psi_a$, $a=1,2,3$ is a generation index. 

Now the last term in the superpotential~(\ref{newsuper}) 
becomes $\kappa_a \Phi\Psi^2$. (we have taken this couplings 
diagonal, i.e. we work directly in a basis of the RH 
neutrino mass eigenstates). Let us assume for simplicity, 
that the neutrino Dirac terms are also diagonal in this 
basis: $g_a l_a\Psi_a\varphi_2$. In fact, there are 
the fermion mass models of this type (see e.g. Ref. 
\cite{rossi}), in which the neutrino mass matrix is diagonal 
and non-zero mixing angles in lepton sector 
emerge exclusively from the charged lepton mass matrix. 

Let us assume that all three constants 
$\kappa_{1,2,3}$ are enough large, namely 
$\kappa_a > \lambda$, in order to evade the excessively  
large reheating temperature of the Universe.   
In this case, the hierarchy of neutrino masses goes as 
$m_a \equiv m_{\nu a} \propto g_a^2/\kappa_a$,  
and it should emerge from the hierarchy 
of the constants $g_{1,2,3}$.

The evolution pattern of the classic fields can be extended 
for the case of three RH sneutrinos in a straight forward manner. 
In particular, we see from~(\ref{tr}) that the reheating 
temperature $T_{\rm r}$ is essentially determined by 
the largest neutrino mass. Recalling also that the 
atmospheric neutrino oscillations point to the   
neutrino mass in the range $m_{3} \sim 0.1$ eV, 
we see that $T_{\rm r}\sim 10^9$ GeV can be 
obtained in our model provided that $\eta \sim 10^{-2}$ 
and $\lambda\sim 10^{-2}$, a quite natural parameter 
range in the hybrid model. 
However, the reheating temperature much smaller than this 
estimate is not very appealing since it would need  
unnaturally small value of $\lambda$. 

On the other hand, the amount of produced $B-L$  
crucially depends on the coefficients $x_a = 3H/\Gamma_a$. 
In order to avoid too strong suppression of the 
result~(\ref{final}), at least one of the factors $x_a$ 
should be larger than 1 or so. 
Interestingly, this suppression factor is inverse proportional 
to neutrino mass -- indeed, from~(\ref{H}) and (\ref{Gamma})
we see: 
\begin{equation} \label{K}
x_a = \frac{\lambda}{\kappa_a^2} 
\left(\frac{10^{-4} ~{\rm eV}}{m_a}\right)
\end{equation} 
and thus the the largest contribution to $B-L$ is given by the 
lightest neutrino mass, presumably $\nu_1$. 
Thus,  the condition $x_1 > 1$ implies the upper bound on 
the lightest neutrino mass 
$m_{1} < (\lambda/\kappa_1^2)\times 10^{-4} ~ {\rm eV}$, 
which limit e.g. for $\lambda < 10^{-2}$ and 
$\kappa_1 >  0.1$, leads to $m_{1} < 10^{-4}$ eV.   
This limit can be naturally met by the by the mass of 
the first generation neutrino, if the neutrino mass hierarchy 
is about the same as that of charged leptons \cite{rossi}.  

Let us conclude by summarising some interesting features
of our model, which is just the simple supersymmetric 
hybrid model with the superpotential~(\ref{newsuper}). 
The couplings $\kappa_a\Phi\Psi_a^2$ of the auxiliary 
orthogonal superfield $\Phi$ to the RH neutrino ones 
$\Psi$ puts the bridge between the inflation and particle 
physics sectors, thus connecting the inflation scale 
$\sim 10^{15}$ GeV to the RH neutrino mass scale needed 
in the context of seesaw mechanism. 
As a bonus, these terms can help in solving many 
problems of the inflationary cosmology and baryogenesis. 

First of all, they allow the orthogonal field oscillations 
to decay enough fastly and thus can prepare the proper 
initial conditions for the inflation onset starting 
from almost arbitrary initial field configurations with the 
classical field values order $M_{\rm P}$. 

And second, at the epoch of postinflationary field 
oscillations, these terms generate the dynamical lepton 
number breaking for the RH sneutrino fields after the 
end of inflation and before the end of reheating era.   
During de-Sitter era these fields are intrinsically 
massless modes and evolve very slowly due to quartic 
self-couplings or because of order $H$ mass term induced 
by the supergravity corrections. 
In either way, at the end of the inflation these fields 
have non-zero values order $H$. 
This is a virtue of ${\rm R}$-symmetry which actually forbids 
terms like  $S\Psi^2$ or $M\Psi^2$ in the superpotential 
~(\ref{newsuper}). 
After inflation they start to oscillate near origin and 
produce the $B-L$ asymmetry of the Universe. 
This happens in an elegant way because the associated $U(1)$ 
charge is dynamically broken in the background of the 
oscillating inflaton fields. After the RH neutrino decay, 
the produced $B-L$ number density is transfered 
to the standard particles, and being reprocessed 
by sphalerons, gives rise to the net baryon 
asymmetry of the Universe. 
In difference from the usual leptogenesis mechanisms  
with the RH neutrino decay 
\cite{buchmuller,fukugita,giudice}, 
our mechanism does not require the presence of 
CP-violation in the lepton mixing. 
So, it can work in the context of predictive models 
\cite{rossi} which do not contain these CP-phases but are  
appealing in all other respects. 

In our model interesting relations emerge between 
the inflationary parameters, reheating temperature, 
$B-L$ number density and neutrino masses. 
The amount of the produced $B-L$ solely depends upon the 
reheat temperature $T_{\rm r}$ with some coefficient 
$X$ which incorporates the coupling and can be order 1. 
In this case, the correct amount of the $B-L$ is obtained 
when $T_{\rm r}\sim 10^9$ GeV, close to the upper bound 
from the thermal gravitino production. 
On the other hand, the possibility of factor $X$ to be 
order 1, implies the upper limit on the lightest neutrino 
mass (presumably $\nu_e$).  
In general, our model is compatible also 
with the neutrino mass spectrum inferred from the 
atmospheric and solar neutrino oscillations. 
One could envisage, that in the context of our observation, 
the hybrid models \cite{mirror} 
designed for explaining the reheating temperature 
difference between the ordinary and hidden (mirror or shadow) 
worlds, could also generate the non-zero baryon asymmetry 
in both sectors. 

Concluding, our model does three jobs very neatly.
First, it correctly indicates the neutrino mass range 
by linking the inflation scale to the RH neutrino mass 
scale in the context of seesaw mechanism. 
Second, it provides dynamically the proper initial condition 
for the onset of inflation in hybrid model. 
And finally, via the classic RH sneutrino fields, 
it generates the proper baryon asymmetry of the universe  
at the epoch of post-inflationary oscillations and reheating.

%%%%%%%%%%%%%%%%%%%%%%%%%%%%%%%%%%%%%%%%%%%%%%%%%%%%%%%%%%%%%%%%%%%%%%%%%%     
\acknowledgements
We thank R. Allahverdi, M. Bastero-Gil, D. Comelli, A. Dolgov, K. Enqvist
and K. Heitmann for valuable discussions. A.M. acknowledges the support of 
{\it The Early Universe network} HPRN-CT-2000-00152. The work of Z.B. is 
partially supported by the MURST research grant {\it Astroparticle Physics}.  

%%%%%%%%%%%%%%%%%%%%%%%%%%%%%%%%%%%%%%%%%%%%%%%%%%%%%%%%%%%%%%%%%%


\begin{references}

\bibitem{linde0} A.D. Linde, Phys. Lett. B259 (1991) 38; 
Phys. Rev. D 49 (1994) 748.

\bibitem{liddle} 
For a review, see 
D.H. Lyth, A.Riotto, Phys. Rep. 314 (1999) 1;  
A. Liddle, D. Lyth, {\it Cosmological Inflation and 
Large-Scale Structure}, Cambridge University Press, 2000.

\bibitem{susy1} 
G. Dvali, Q. Shafi, R. Schaefer, Phys. Rev. Lett. 73 (1994) 1886;
E. Copeland et al., Phys. Rev. D 49 (1994) 6410;  
E. Stewart, Phys. Rev. D 51 (1995) 6847; 
G. lazarides, C. Panagiotakopoulos, Phys. Rev. D 52 (1995) 559.  

\bibitem{susy2} 
P. Bin\'etruy, G. Dvali, Phys. Lett. B388 (1996) 241; 
E. Halyo, Phys. Lett. B387 (1996) 43.  

\bibitem{piran}
D.S. Goldwirth, T. Piran, Phys. Rev. Lett. 64 (1990) 2852; 
T. Vachaspati, M. Trodden, Phys. Rev. D 61 (2000) 023502.

\bibitem{tetradis} 
G. lazarides, C. Panagiotakopoulos, N.D. Vlachos, 
Phys. Rev. D 54 (1996) 1369; 
G. lazarides, N.D. Vlachos, Phys. Rev. D 56 (1997) 4562; 
N. tetradis, Phys. Rev. D 57 (1998) 5997;  
G. Lazarides, N. Tetradis, Phys. Rev. D 58 (1998) 123502. 
  
\bibitem{zurab} 
Z. Berezhiani, D. Comelli, N. Tetradis, 
Phys. Lett. B431 (1998) 286. 

\bibitem{sakh}
A.D. Sakharov, JETP Lett. B 91 (1967) 24.

\bibitem{buchmuller}
A. Dolgov, Phys. Rep. 222 (1992) 309; 
W. Buchmuller, M. Plumacher, Phys. Rep. 320 (1999) 329;
A. Riotto, M. Trodden, Ann. Rev. Nucl. Part. Sci.  49 (1999) 35.

\bibitem{fukugita} 
M. Fukugita, T. Yanagida, Phys. Lett. B174 (1986) 45; 
M.A. Lutty, Phys. Rev. D 45 (1992) 455.

\bibitem{thooft} 
G. t'Hooft, Phys. Rev. Lett. 37 (1976) 8;
V. Kuzmin, V. Rubakov, M. Shaposhnikov, Phys. Lett. B155 (1985) 36; 
J. Ambjorn et al., 
Nucl. Phys. B353 (1991) 346.

\bibitem{affleck}
I. Affleck, M. Dine, Nucl. Phys. B249 (1985) 361.

\bibitem{seesaw}
M. Gell-Mann, P. Ramond, R. Slansky, in {\it Supergravity},
eds. P. van Niewenhuizen and D.Z. Freedman (North Holland 1979);
T. Yanagida, Proceedings of {\it Workshop on
Unified Theory and Baryon number in the Universe}, eds.
O. Sawada and A. Sugamoto (KEK 1979);
R.N. Mohapatra, G. Senjanovi{\'c}, Phys. Rev. Lett. {\bf 44}, 912
(1980).

\bibitem{mar1}
M. Bastero-Gil, S.F. King, and J. Sanderson, 
Phys. Rev. D 60 (1999) 103517; 

\bibitem{mar2}
M. Bastero-Gil, and A. Mazumdar, Phys. Rev. D 62 (2000) 083510. 

\bibitem{dan} 
D. Cormier, K. Heitmann, and A. Mazumdar, hep-ph/0105236.

\bibitem{sarkar} 
For a review, see S. Sarkar, Rep. Prog. Phys. 59 (1996) 1493. 

\bibitem{kofman} 
A.D. Dolgov, D. Kirilova, Yad. Fiz. 51 (1989) 273 
[Sov.J.Nucl.Phys. 50 (1989) 172];
J. Traschen, R. Brandenberger, Phys. Rev. D 42 (1990) 2491;
L. Kofman, A. Linde, A. Starobinsky, 
Phys. Rev. Lett. 73 (1994) 3195;
Y. Shtanov, J. Traschen, R. Brandenberger, 
Phys. Rev. D 51 (1995) 5438;
L. Kofman, A. Linde, A. Starobinsky, 
Phys. Rev. D 56 (1997) 3528; 
J. Baacke, K. Heitmann, C. Patzold, Phys. Rev. D 58 (1998) 125013;
P.B. Greene, L. Kofman, Phys. Lett. B448 (1999); 
A.L. Maroto, A. Mazumdar, Phys. Rev. D 59 (1999) 08350. 

\bibitem{maroto}
A.L. Maroto and A. Mazumdar, Phys. Rev. Lett. 84 (2000) 1655;

\bibitem{grav} 
G.F. Giudice, I. Tkachev, A. Riotto, JHEP 9908 (1999) 009; 
R. Kallosh, L. Kofman, A. Linde, A. Van Proyen, 
Phys. Rev. D 61 (2000) 103503; 
 
\bibitem{ruz}
R. Allahverdi, M. Bastero-Gil, A. Mazumdar, 
Phys. Rev. D 64 (2001) 023516.

\bibitem{giudice}
G.F. Giudice et al., JHEP 9908 (1999) 014.  

\bibitem{felder}
G. Felder et al., hep-ph/0012142.

\bibitem{kleb} 
S. Khlebnikov, M.E. Shaposhnikov, Nucl. Phys. B308 (1998) 885;
J.A. Harvey, M.S. Turner, Phys. Rev. D 42 (1990) 3344.

\bibitem{rossi}
Z. Berezhiani, A. Rossi, Nucl. Phys. B594 (2001) 113; 
JHEP 9903 (1999) 002; For earlier models, see 
Z. Berezhiani, J. Chkareuli, JETP Lett. 37 (1983) 338; 
Z. Berezhiani, Phys. Lett. B150 (1985) 177.

\bibitem{mirror} 
Z. Berezhiani, A. Dolgov, R.N. Mohapatra, 
Phys. Lett. B375 (1996) 26; 
Z. Berezhiani, D. Comelli, F. Villante, 
Phys. Lett. B503 (2001) 362.  

\end{references}
\end{document}